\documentclass{osa-article}
\pdfoutput=1
\journal{osajournal}

\articletype{Research Article}
\usepackage{graphicx}
\usepackage{xcolor}
\begin{document}
\title{A Prototype of a Compact Rubidium-Based Optical Frequency Reference for Operation on Nanosatellites}
\author{Aaron Strangfeld\authormark{1,2}, Simon Kanthak\authormark{1,2}, Max Schiemangk\authormark{2}, Benjamin Wiegand\authormark{1}, Andreas Wicht\authormark{2}, Alexander Ling\authormark{3,4} and Markus Krutzik\authormark{1,2}}
\date{}
\address{\authormark{1}Department of Physics, Humboldt-Universität zu Berlin, Newtonstraße 15, 12489 Berlin, Germany\\
\authormark{2}Ferdinand-Braun-Institut, Leibniz-Institut für Höchstfrequenztechnik, Gustav-Kirchhoff-Strasse 4. 12489 Berlin, Germany\\
\authormark{3}Centre for Quantum Technologies, National University of Singapore, Block S15, 3 Science Drive 2, Singapore 117543\\
\authormark{4}Department of Physics, National University of Singapore, Block S12, 2 Science Drive 3, Singapore 117551}




\begin{abstract}
Space-borne optical frequency references based on spectroscopy of atomic vapors may serve as an integral part of compact optical atomic clocks, which can advance global navigation systems, or can be utilized for earth observation missions as part of laser systems for cold atom gradiometers. Nanosatellites offer low launch-costs, multiple deployment opportunities and short payload development cycles, enabling rapid maturation of optical frequency references and underlying key technologies in space. Towards an in-orbit demonstration on such a platform, we have developed a CubeSat-compatible prototype of an optical frequency reference based on the D2-transition in rubidium. A frequency instability of 1.7$\times$10$^{-12}$  at 1\,s averaging time is achieved. The optical module occupies a volume of 35\,cm$^3$, weighs 73\,g and consumes 780\,mW of power.
\end{abstract}

\pagenumbering{gobble} 
\section{Introduction}
Optical frequency references (OFR) based on spectroscopy of atomic vapors can be used as key components of optical clocks and quantum sensors based on cold atoms. Prospective applications in space range from time keeping \cite{SpaceClocks}, navigation \cite{DeepSpaceNav}, geodesy \cite{Geodesy} to high-precision measurements in fundamental physics, such as clock comparisons \cite{Pharao} and gravitational wave detection \cite{AEDGE}. As OFRs are on their way from realizations in the laboratory to field applications, further development efforts and increased technical maturity are required, especially in case of cost-intensive space missions that demand a low risk of failure for all subsystems.\\ CubeSats, nanosatellites made up of units of about 1 kg in mass and $10\times10\times10$\,cm$^3$ in volume, offer a platform for cost-effective and fast-paced development and testing, incrementally increasing the technology readiness of the payloads \cite{CubeSatsforQuantum}. Apart from building up space heritage, this in-orbit qualification can also close gaps between laboratory setups and potential commercial applications. Such efforts have already been carried out successfully for quantum communication by demonstrating in-orbit quantum entanglement on-board a CubeSat \cite{SpooQy}. A mission for orbit-to-ground transmission of entanglement on this platform \cite{ROKS} is scheduled for 2022 and a CubeSat containing a cold atom system \cite{CASPA} has been presented, recently.\\
OFRs based on optical lattices or trapped ions currently offer the highest achievable performance with fractional frequency instabilities at the $10^{-18}$ level at timescales greater than $10^3$\,s \cite{LatticeClock,IonClock}. Systems based on atomic beams and Ramsey-Bordé interferometry can be smaller while still achieving fractional frequency instabilities at the order of $10^{-15}$ at timescales greater than 1\,s\cite{BeamClock}. However, even compact versions of both types \cite{CompactLatticeClock, CompactAtomicBeam} exceed the size, weight and power (SWaP) budget of CubeSats. In contrast, OFRs based on atomic vapors have the potential to meet these SWaP requirements. The most compact and therefore easiest to integrate devices are chip-scale atomic clocks (CSAC), which make use of coherent population trapping (CPT) \cite{CPT}, but only achieve instabilities of $10^{-10}$ at 1\,s averaging time \cite{CSAC2004,CSAC2008, CSAC2018}. OFRs based on Doppler-free spectroscopy of alkali metals are more complex, but compact designs can achieve instabilities 1000 times lower. The operation of this type of atomic vapor based OFRs in space has already been shown during various sounding rocket missions \cite{Kalexus, Fokus, Jokarus} and the Cold Atom Laboratory mission onboard the International Space Station \cite{CAL}.  Miniaturization is realized by using small optical breadboards \cite{RbD22006,Gain,RbD22017,2Photon2018} or micro-integration of optical components \cite{KD22019,CsD22020} around the alkali vapor cells. Especially, integration of the light source and the spectroscopy unit into one package \cite{RbD22019} as well as the use of MEMS vapor cells \cite{2photon,MEMSD2,2PhotonandComb} enable major size reduction of the systems. Reaching the SWaP threshold for nanosatellite implementation is thus realistic.\\
In this work, we present a miniaturized prototype of an OFR for operation on a CubeSat, which is based on the spectroscopy of the D2-transition in rubidium using a micro-integrated distributed feedback (DFB) laser diode. The paper is structured as follows. First, the design of the module is presented in Section\,\ref{sec:design}. Second, the operating conditions are described in Section\,\ref{sec:operation}. Third, the performance evaluation is shown in Section\,\ref{sec:performance}, where the Allan deviation of the frequency  is determined by using an additional reference stabilized to the D2-transition.  Finally, an outlook towards a complete CubeSat payload is given in Section\,\ref{sec:outlook}.

\section{Design}\label{sec:design}
\begin{figure}[htbp]
\centering
\includegraphics[width=0.7\linewidth,trim={0 10cm 0 0},clip]{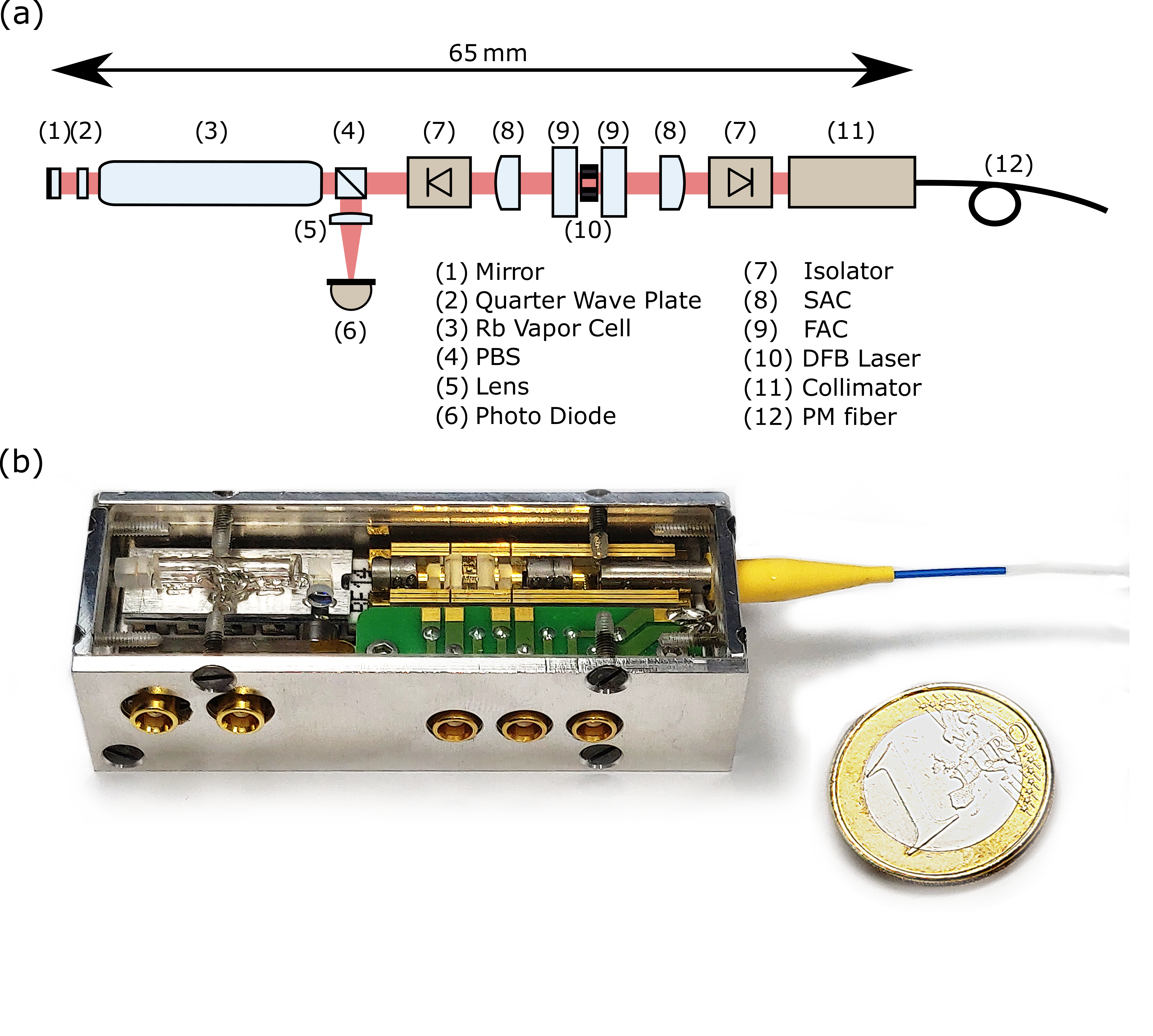}
\caption{Design of the OFR. (a) The optical components are arranged longitudinally. This leads to simple alignment but also to the vapor cell being close to the rear optical isolator. The sizes of the components are true to scale. (b) The assembled module is shown with a One-Euro coin for scale.  The electrical feedthroughs are from left to right: detector power, detector signal, DFB current, NTC and TEC. }
\label{fig:Optics Design}
\end{figure}

\begin{figure}[htbp]
\centering
\includegraphics[width=0.65\linewidth]{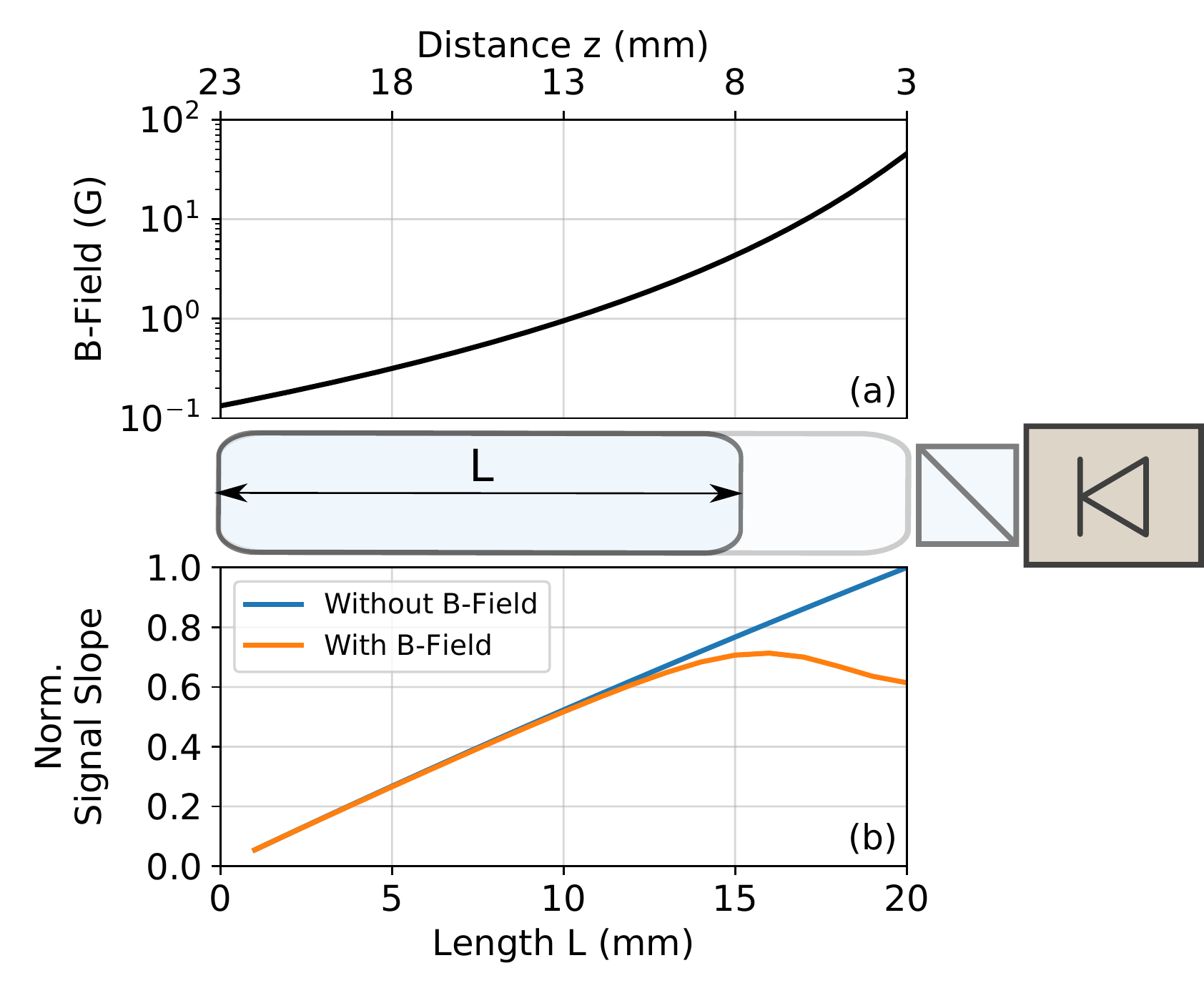}
\caption{Simulation of the effect of the rear isolator's magnetic field on the slope of the FMS signal. (a) Distance-dependent magnetic field of the semi-double stage isolator with the distance $z$ measured from its edge. (b) Simulated signal slope for different cell lengths normalized to its value with a cell length of 20\,mm and no magnetic field. The schematic between the graphs visualizes the cell length variation at maximum distance to the optical isolator.}
\label{fig:Simulations}
\end{figure}

The optical design of the OFR is realized by a linear arrangement of the optical components as depicted in Figure\,\ref{fig:Optics Design}a and is inspired by a previously developed master oscillator power amplifier system \cite{Mopa}.
A pair of lenses (Fast Axis Collimating (FAC) and Slow Axis Collimating (SAC)) collimates the light on each side of a 1.5\,mm long GaAs DFB laser \cite{FBHDFB}. Semi-double stage optical isolators (60\,dB) protect the laser diode against optical feedback. The light emitted from the laser's front (right) output is coupled into a polarization maintaining (PM) single-mode fiber. The light emitted from the laser's rear (left) output is used for spectroscopy of rubidium utilizing a vapor cell made of borosilicate glass with angled windows. The rear facet emits only about 3\,\% of the total optical output power due to its reflective coating. At the working point, this corresponds to 2\,mW directly from the diode. Proper attenuation is achieved by rotating the rear optical isolator. A beam diameter of 600\,\textmu m and an optical power incident on the vapor cell of 30\,\textmu W correspond to a saturation parameter of $S=4$. Based on preceding breadboard system investigations, we expected reaching the optimum slope-to-noise ratio at this level of saturation. The retro-reflecting scheme, where the pump beam is turned into the probe beam at the mirror, enables Doppler-free spectroscopy \cite{DopplerFree}. A double-passed quarter wave plate between the vapor cell and the aforementioned mirror rotates the polarization axis such that the reflected probe beam can be deflected by the polarizing beam splitter (PBS). A lens focuses the probe light onto a fast photo diode with a bandwidth of 500\,MHz, which is mounted on an amplifying circuit board. Frequency modulation spectroscopy (FMS) \cite{FMS}, realized by modulation of the laser's injection current at a frequency of $6.6$\,MHz, generates an error signal for a feedback-loop. The injection current into the laser also serves as the actuator for the stabilization of the laser's emission frequency.\\
The optical components are mounted on 1\,mm thick benches made of aluminum for the spectroscopy unit on the left and aluminum nitride (AlN) for the laser unit on the right in Figure\,\ref{fig:Optics Design}b. The coefficient of thermal expansion of AlN is close to the one of the GaAs DFB laser chip. This leads to reduced mechanical stress on the diode laser under temperature variations. A low-out-gassing, heat conductive adhesive foil connects both benches to a common thermo-electric cooler (TEC) for thermal control. The TEC itself is attached to an aluminum heat sink utilizing the same type of adhesive foil. A 10\,k$\Omega$ negative temperature coefficient (NTC) resistor next to the DFB laser measures the temperature to provide feedback to a temperature controller acting on the TEC. The amplifying detector circuit board and the electronics interface board are attached to the heat sink by screws. The interface board allows for transition from compact electrical connectors to wire bonds for the laser unit. Gold plated and structured AlN rails are used as intermediate platforms for the wire bonds. At the current development stage, the prototype is encased in aluminum walls screwed to the heat sink. Figure\,\ref{fig:Optics Design}b shows the assembled module, named \emph{iQube} (\emph{i}ntegrated \emph{Q}uantum technology subsystem for C\emph{ube}Sats), with an acrylic glass lid for demonstration purposes.\\
The linear design enables simple active alignment of the optical components. Furthermore, the usage of a single TEC for both the laser diode and the vapor cell reduces the number of necessary temperature controllers. The size constraints of a 1U CubeSat and the space required for fiber bending limit the module's length to 70\,mm. Table\,\ref{tab:swap} summarizes iQube's SWaP as it is depicted in Figure\,\ref{fig:Optics Design}b. We achieved a coupling efficiency of 37\,\%  corresponding to an optical output power of 8\,mW with a minimum polarization extinction ratio of 31\,dB. Higher coupling efficiency could have been achieved by optimization of the position of both collimating lenses simultanously. We optimized the position and attached the lenses one after the other as we were limited by the used assembly facilities.
\begin{table}[htbp]
\centering
\caption{\bf Size, Weight and Power consumption of the optical module as depicted in Figure \ref{fig:Optics Design}b.}
\begin{tabular}{ll}
\hline
Size & $70\times 26\times 19.2$\,mm$^3$\\
Weight & $73$\,g\\
Power & $780$\,mW\\
\hline
\end{tabular}
  \label{tab:swap}
\end{table}
\\
Due to the constraints regarding the length of the module, we had to place the rear optical isolator and the vapor cell within 28\,mm. With a length of 5\,mm and 2.5\,mm of the optical isolator and the PBS, respectively, 20\,mm remained for the vapor cell if a minimum distance of 0.25\,mm is kept between the components. To quantitatively evaluate the effect of the isolator's magnetic field, its impact for different cell lengths was simulated.\\The isolator (Isowave I-780-LM-SD-1.4-4) contains two oppositely oriented samarium cobalt (SmCo28) ring magnets with a magnetic remanence of $1.06\times 10^4$\,G. Their centers are separated by a distance of $2.0$\,mm and placed at a distance of $1.5$\,mm to the isolators edge on each side. Based on the geometry of the ring magnets ($0.89$ \,mm length, $1.20$\,mm inner radius and $1.86$\,mm outer radius) the magnetic field could be modelled using the equations for the magnetic field strength of ring magnets (see e.\,g. Appendix F in \cite{MagneticField}).\\Figure\,\ref{fig:Simulations}a shows the resulting distance-dependent magnetic field strength along the optical axis. Based on this field, Doppler-free spectra of the $F=3$ to $F'=4$ transition in $^{85}$Rb were simulated with \emph{ElecSus} \cite{Elecsus1,Elecsus2}, an open source software. The length of the vapor cell was identified with the length of the optical path through a rubidium ensemble. Lorentzians were fitted to the generated spectra and the peak-to-width ratio served as a quantitative measure for the FMS signal slope. The impact of the line splitting depends on the linewidth without magnetic field. We based the simulation on a saturation parameter of $S=4$ resulting in a linewidth of $\gamma_\mathrm{s}=\sqrt{5}\times6$\,MHz\,$=13.4$\,MHz. Figure\,\ref{fig:Simulations}b shows the behaviour of the signal slope with increasing length of the vapor cell (optical path length). Additionally, the slope without the influence of a magnetic field is shown. There is an optimum at a length of 16\,mm. We realized the system with a 15\,mm long vapor cell (17\,mm with windows) placed at a distance of 8\,mm to the optical isolator, which results in 70\,\% of the slope achievable without a magnetic field and maximum cell length.

\section{Operation}
\label{sec:operation}
\begin{figure}[htbp]
\centering
\includegraphics[width=0.7\linewidth]{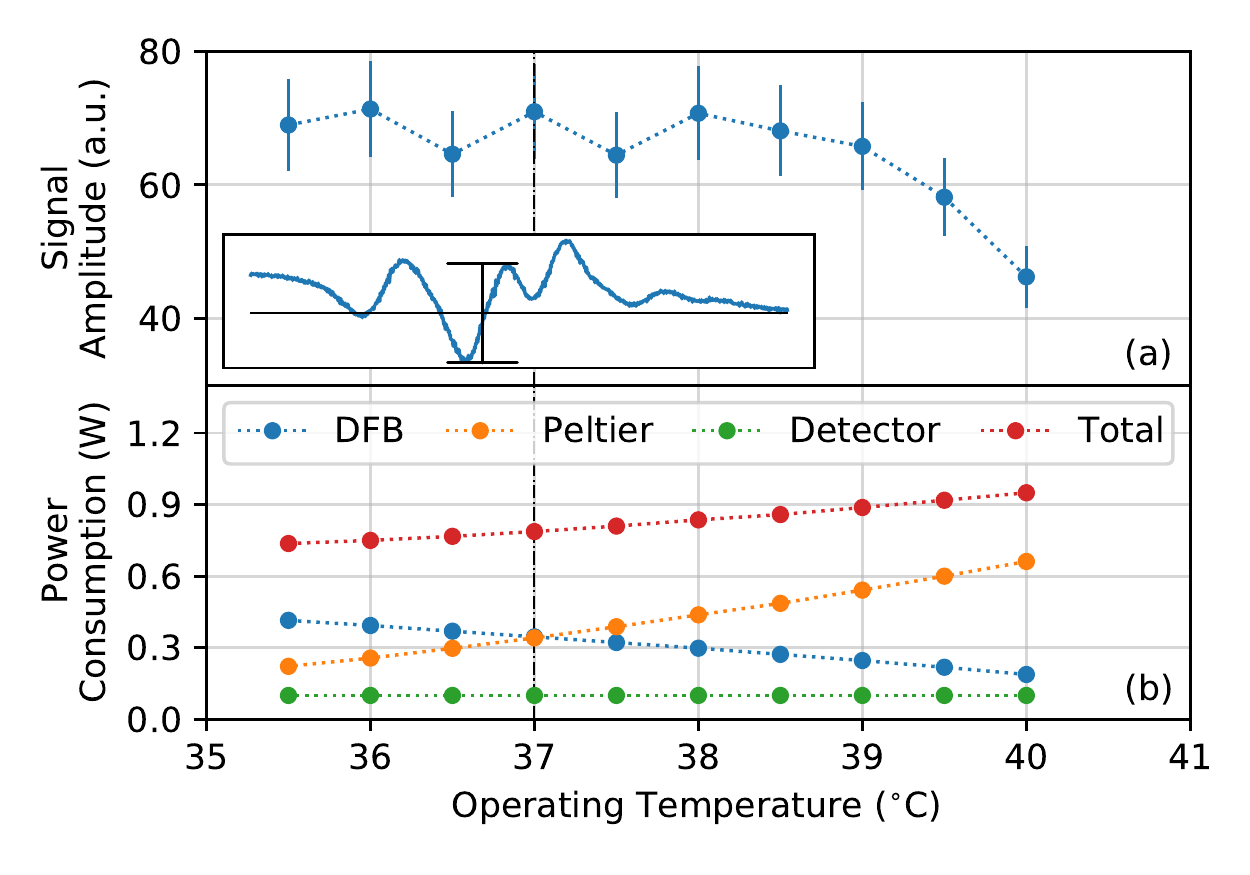}
\caption{Temperature dependent signal amplitude (a) and power consumption of the module under laboratory conditions (b). The inlay in (a) depicts the determination of the amplitude from the spectroscopy signal. With increasing temperature, the driving current of the laser diode is decreased to match the D2-transition's wavelength. The lower optical power results in a signal amplitude decrease at temperatures above 39\,$^{\circ}$C. A working point at 37\,$^{\circ}$C indicated by the dashed gray line was selected. Working points at lower temperatures, higher diode currents and finally lower power consumption would be realizable, but would also further reduce the lifetime of the diode.}
\label{fig:WorkingPoint}
\end{figure}

For stabilization of the laser frequency we developed \emph{Linien} \cite{linien}, an open-source software that runs on a RedPitaya StemLab-14 device. This platform provides fast analog-to-digital and digital-to-analog converters (125 MS/s with a resolution of 14 bit) as well as an on-board field-programmable gate array (FPGA) and a complete Linux system in a compact form factor. On this system, we implemented generation of the modulation signal, demodulation of the spectroscopy signal, digital filtering of the signal and the proportional-integral-differential (PID) controller that generates the control signal  with a servo bandwidth of about $700$\,kHz. In order to automatically optimize the slope of the FMS signal by tuning spectroscopy parameters, (modulation frequency, modulation index and demodulation phase), \emph{Linien} utilizes a covariance matrix adaptation evolution strategy \cite{MachineLearning}. Furthermore, an automatic re-locking algorithm is able to relocate the original locking point after signal loss.\\
A current driver based on a modified Libbrecht-Hall design \cite{LibbrechtHall} supplies the current for the laser diode. Temperature control is realized by a compact commercial controller (Meerstetter TEC-1091).\\
To find an optimum working point under laboratory conditions, the signal amplitude and total power consumption was measured at different temperatures after optimization of the spectroscopy parameters. The temperature measurement is realized by the NTC placed next to the DFB laser. Figure\,\ref{fig:WorkingPoint}a shows that the amplitude enters a plateau at temperatures below 39\,$^{\circ}$C. The FMS signal at 37.0\,$^{\circ}$C is shown with an indication of the determined signal amplitude in the inlay. The decrease at higher temperatures originates from lower optical power from the laser diode as a result of lowered driving current to match the transition's wavelength. Figure\,\ref{fig:WorkingPoint}b shows the temperature dependent power consumption of the subsystems and the total power consumption. The TEC's power consumption is reduced by 70\,\% within a temperature decrease of 4\,K. The total power consumption decreases as well by about 6\,\%/K. A working point of 37.0\,$^{\circ}$C with a total power consumption of 780\,mW is chosen for the performance evaluation. This is a compromise between lower power consumption at lower temperatures and increased lifetime of the laser due to a lower injection current at higher temperatures \cite{LDDegradation}.

\section{Performance}\label{sec:performance}
\begin{figure}[htbp]
\centering
\includegraphics[width=0.7\linewidth]{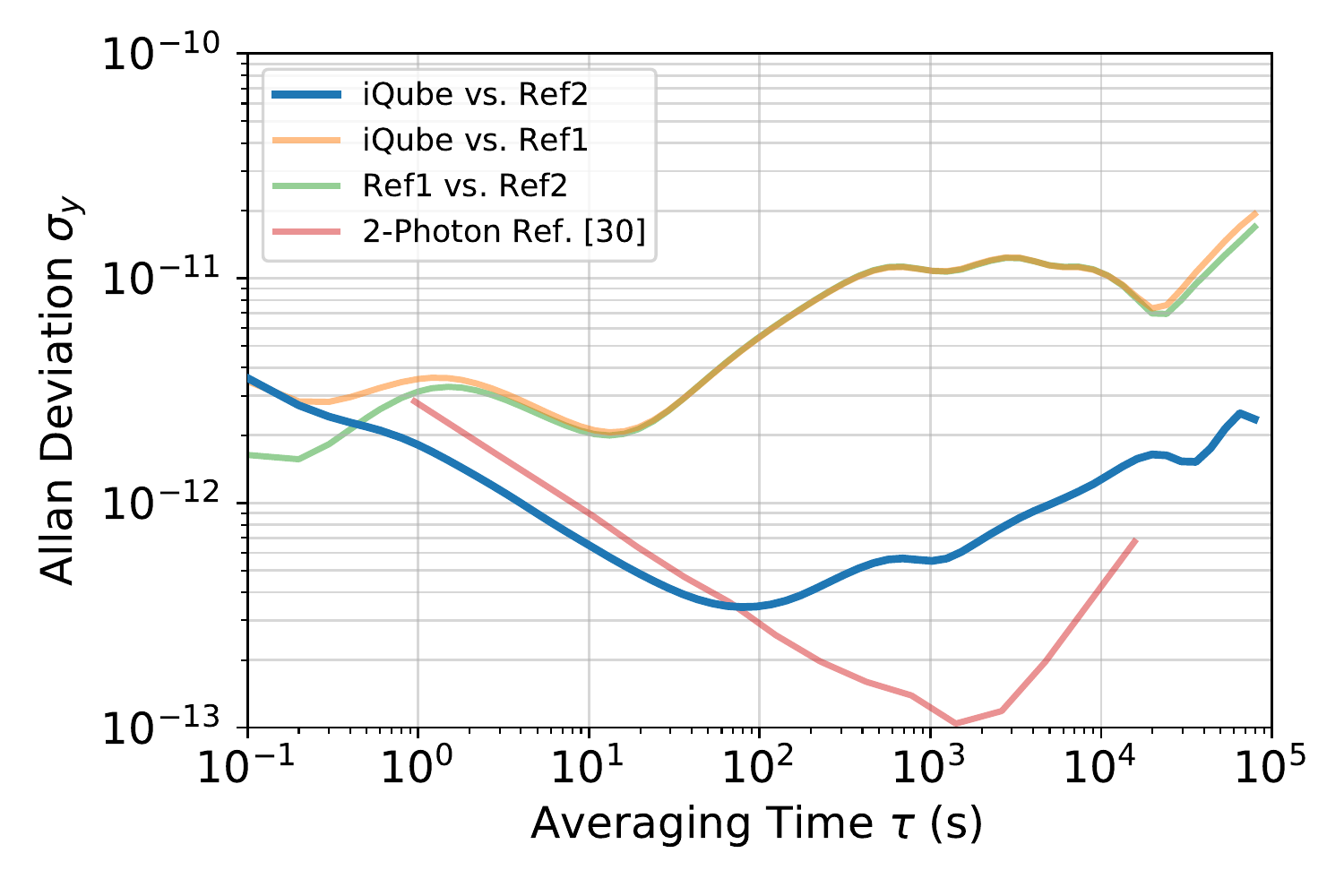}
\caption{Allan deviation of the iQube module determined by beat-note measurements with two external references (Ref1: orange, Ref2: blue). Ref1 has been developed as an FMS-based spectroscopy system for application on sounding rockets \cite{Fokus}. Ref2 is based on MTS and used as a laboratory reference setup. The Allan deviation of a state-of-the-art compact two-photon breadboard system \cite{2photon} is shown for comparison.}
\label{fig:Allan Deviation}
\end{figure}

\begin{figure}[htbp]
\centering
\includegraphics[width=0.7\linewidth]{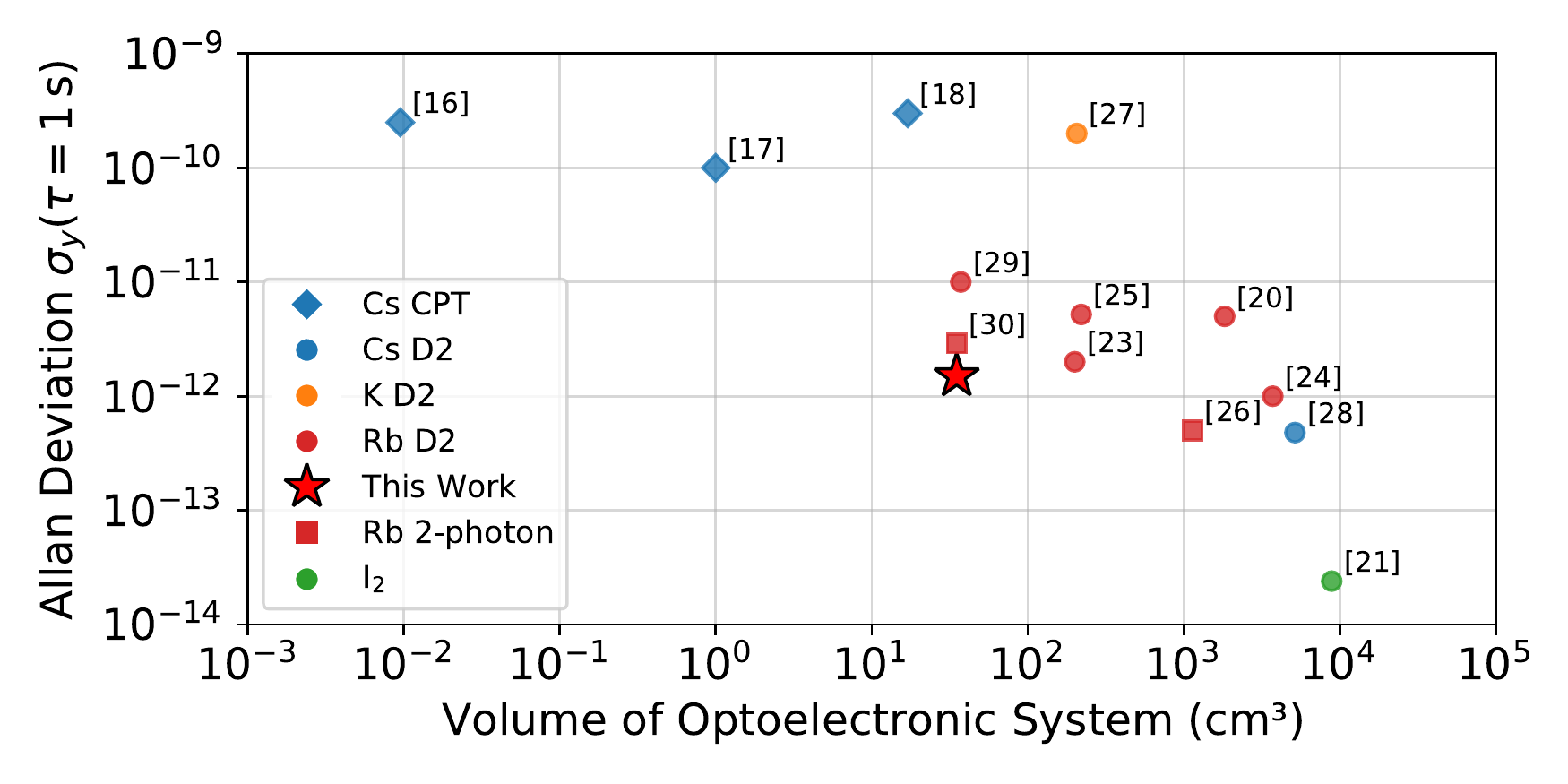}
\caption{Allan deviation at 1\,s averaging time versus the optoelectronic system volume of various atomic vapor based optical frequency references. Shapes and colors indicate the utilized spectroscopy method and atomic species, respectively. All D2-transition based references use Doppler-free spectroscopy. In case of the iodine reference, MTS of a molecular transition is used. To ensure comparability, the volume of control electronics was not taken into account. Allan deviations of the OFRs used in \cite{Fokus} and \cite{Gain} have been evaluated again in this work and in case of \cite{RbD22019}, the Allan deviation was estimated based on a power spectral density.}
\label{fig:DataOptRef}
\end{figure}

The performance of the module was evaluated in terms of the Allan deviation of the relative frequency deviation $y=\frac{\Delta\nu}{\nu_0}$:
\begin{equation}
\sigma_y(\tau) = \sqrt{\frac{1}{2}\langle (y_{n+1}-y_{n})^2\rangle},
\end{equation}
where $n$ is the index of the measured time series of length $\tau$. The frequency deviation was measured by beat-note measurements, where two additional frequency references "Ref1" \cite{Fokus} and "Ref2" \cite{Gain} stabilized on the D2-transition of rubidium were used.  Ref1 was previously developed for technology demonstration on a TEXUS sounding rocket mission and is based on frequency modulation of the laser current. Ref2 serves as a reference laser for an atom interferometer and is based on modulation transfer spectroscopy (MTS) \cite{MTS}. In each reference a DFB laser is used. The resulting Allan deviation is an upper limit for iQube's contribution to the frequency instability. \\\\
Figure\,\ref{fig:Allan Deviation} shows the  Allan deviations of the three beat-note frequencies. The presented module achieves an Allan deviation of $1.7\times10^{-12}$ at 1\,s averaging time and a minimum of $3.4\times10^{-13}$ at $10^2$\,s based on the beat-note with Ref2. At times larger than $10^2$\,s there is an increase of the Allan deviation which cannot be compensated by removing a linear drift. We expect this to be the result of ambient temperature variations affecting iQube or Ref2.\\
The frequency stability is also compared to the one achieved  by a compact breadboard system for two-photon spectroscopy of rubidium presented in \cite{2photon}. There, a distributed Bragg reflector laser is used.
Figure\,\ref{fig:DataOptRef} shows a comparison of iQube's performance, measured by the Allan deviation at 1\,s averaging time, and volume to other compact optical frequency references based on atomic vapors. Our module lines up with the other most compact systems that achieve Allan deviation of $10^{-11}$ and below with 1\,s averaging time: \cite{RbD22019} and \cite{2photon}.
\section{Discussion \& Outlook}\label{sec:outlook}
\begin{figure}[htbp]
\centering
\includegraphics[width=0.5\linewidth]{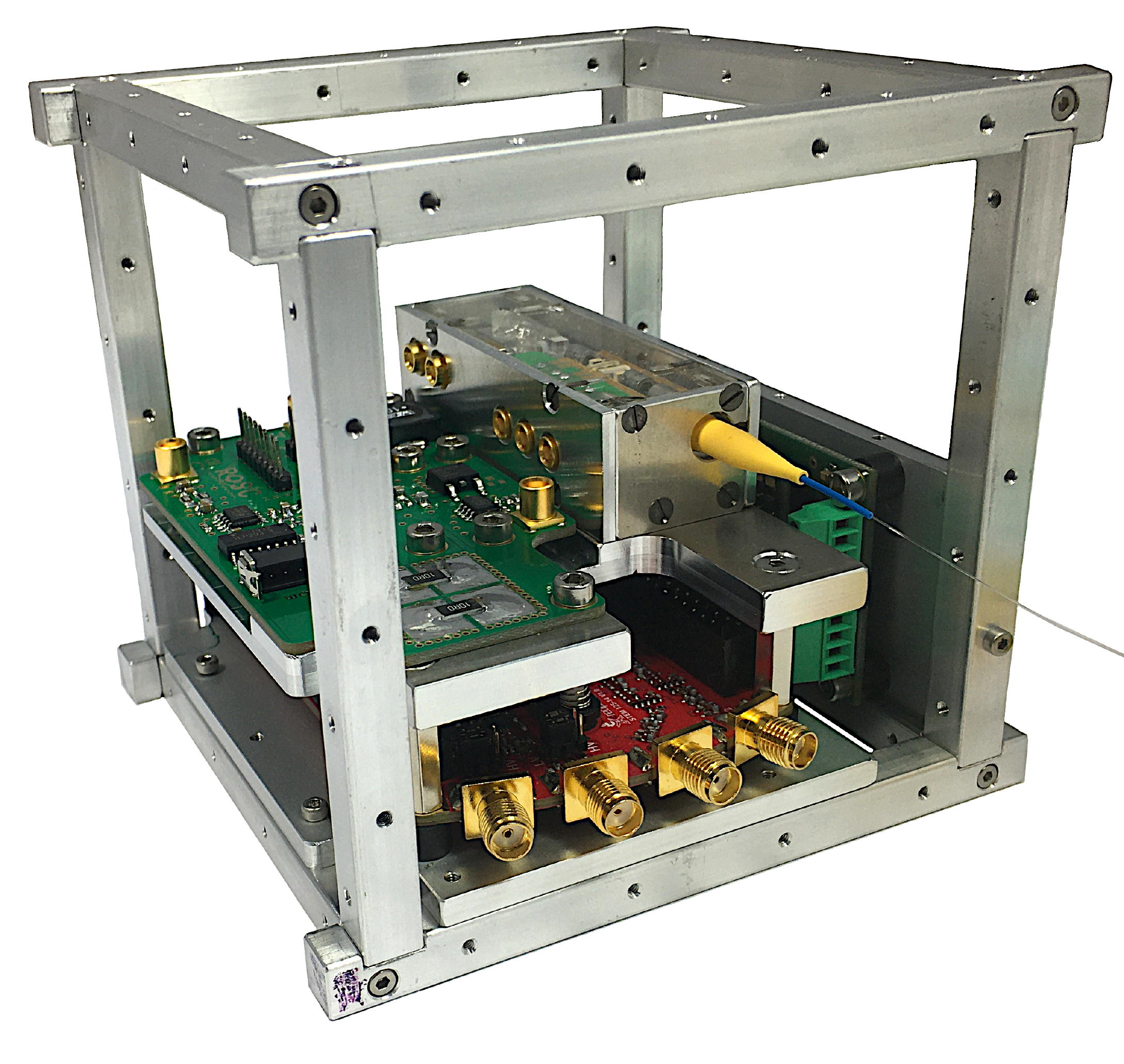}
\caption{Assembled demonstrator system with complete functionality of the OFR inside a volume of 0.6U of a Cubesat with the following components: (bottom) RedPitaya StemLab for signal demodulation and modulation as well as PID-control of the laser current, (right) Meerstetter TEC-1091 temperature controller, (upper left) current driver and (upper right) iQube OFR.}
\label{fig:CubesatDemonstrator}
\end{figure}

We have presented the design and performance evaluation of a compact fiber-coupled OFR based on spectroscopy of the D2-transition in rubidium. With a volume of $70\times 26\times 19.2$\,mm$^3$, the module fits into a 1U CubeSat and occupies 3.5\,\% of its volume. The mass of 73\,g consumes 5\,\% of the mass budget. The evaluation of the power consumption of 780 mW needs to be done in relation to the complete payload including a fast processing system and a temperature and current controller. Figure\,\ref{fig:CubesatDemonstrator} shows such a system occupying 0.6U of a CubeSat. The RedPitaya StemLab equipped with an FPGA is at the bottom, the temperature controller is on the right, the current driver is in the upper left and the optical module is placed in the upper right. The power consumption of this payload would be 6.4\,W under laboratory conditions, which is dominated by the RedPitaya StemLab consuming 4.0\,W when idle and 4.8\,W when operating the OFR. A power budget of this size could only be realized by deployable solar panels on a 1U CubeSat, which would, however, increase the risk of failure during satellite deployment. A realization on a 2U or 3U CubeSat with a higher number of solar panels attached to its hull would thus be more realistic. A presented power consumption of 12\,W of a complete operating cold atom CubeSat system \cite{CASPA} hints towards the possibility of reduced power consumption especially for the control electronics.\\
At an averaging time of 1\,s, the iQube OFR achieved an Allan deviation of $1.7\times10^{-12}$. With this performance, the frequency reference is well suited to be used in laser systems for cold atoms experiments or optical metrology applications such as wavelength meter calibration \cite{highfinesse}. It also represents a technology demonstrator for future compact vapor cell based optical clocks in space. Adapting the module to a fluorescence based two-photon \cite{2photonWithDFB} or an absorption based 5S-6P transition scheme \cite{420scheme} in rubidium could be an option for higher stability with similar SWaP.\\
Necessary efforts towards the realization of a mission-ready system include technical improvements, development of interfaces to the CubeSat platform systems and execution of the qualification process (vibration, thermal vacuum, out-gassing, radiation). On the technical side, hermetic sealing of the optical module is required to protect the laser diode. The size of the electrical interface can be reduced in this step, as well. Further size reduction could also be achieved by merging the current driver and the temperature controller into a custom laser controller and a single space-qualified and more efficient on-board computer with an FPGA could replace the RedPitaya by controlling both, the platform systems and the payload. A simple mission scenario for technology demonstration would then consist of two OFR systems and a detector connected to a frequency counter to record the optical beat note frequency.

\section*{Funding}
\section*{Acknowledgments}
This work has been done in a joint collaboration between Humboldt-Universität zu Berlin and National University of Singapore, supported by the Berlin University Alliance. \\\\
The micro-integration was realized at Ferdinand-Braun-Institut, Leibniz-Institut für Höchstfrequenztechnik.\\\\
A. Strangfeld, S. Kanthak, B. Wiegand and M. Krutzik additionally acknowledge support by the German Space Agency DLR with funds provided by the Federal Ministry for Economic Affairs and Energy (BMWi) under grant numbers 50RK1971 (ROSC), 50WM2066 (OPTIMAL-QT) and 50RK1978 (QCHIP).\\\\
The authors greatly thank M. Christ and A. Stiekel for the preparation and provision of the optical assembly facilities, O. Anton and B. Leykauf for providing the optical frequency references for testing and the final measurement, K. Döringshoff for advice during the performance evaluation, M. Schoch and C. Kaiser for technical support and R. Kumar and S. Tapiawala for fruitful discussions.
\section*{Disclosures}
The authors declare no conflicts of interest.
\bigskip
\bibliography{sample.bib}
\end{document}